\documentclass[journal=jacsat,manuscript=article]{achemso}

\usepackage[version=3]{mhchem} 

\author{Bing Chen}
\affiliation{School of Electronic Science and Applied Physics,Hefei University of Technology, Hefei, Anhui 230009, China}

\author{Xianfei Hou}
\affiliation{School of Electronic Science and Applied Physics,Hefei University of Technology, Hefei, Anhui 230009, China}

\author{Feifei Ge}
\affiliation{School of Electronic Science and Applied Physics,Hefei University of Technology, Hefei, Anhui 230009, China}

\author{Xiaohan Zhang}
\affiliation{School of Electronic Science and Applied Physics,Hefei University of Technology, Hefei, Anhui 230009, China}

\author{Yunlan Ji}
\affiliation{School of Electronic Science and Applied Physics,Hefei University of Technology, Hefei, Anhui 230009, China}

\author{Hongju Li}
\affiliation{School of Electronic Science and Applied Physics,Hefei University of Technology, Hefei, Anhui 230009, China}

\author{Peng Qian}
\affiliation{School of Electronic Science and Applied Physics,Hefei University of Technology, Hefei, Anhui 230009, China}

\author{Ya Wang}
\email{ywustc@ustc.edu.cn}
\affiliation{Hefei National Laboratory for Physical Sciences at the Microscale and Department of Modern Physics,
University of Science and Technology of China, Hefei 230026, China}
\alsoaffiliation{CAS Key Laboratory of Microscale Magnetic Resonance, USTC}
\alsoaffiliation{Synergetic Innovation Center of Quantum Information and Quantum Physics, USTC}

\author{Nanyang Xu}
\email{nyxu@hfut.edu.cn}
\phone{+0551 62902751}
\fax{+0551 62919106}
\affiliation{School of Electronic Science and Applied Physics,Hefei University of Technology, Hefei, Anhui 230009, China}

\author{Jiangfeng Du}
\email{djf@ustc.edu.cn}
\affiliation{Hefei National Laboratory for Physical Sciences at the Microscale and Department of Modern Physics,
University of Science and Technology of China, Hefei 230026, China}
\alsoaffiliation{CAS Key Laboratory of Microscale Magnetic Resonance, USTC}
\alsoaffiliation{Synergetic Innovation Center of Quantum Information and Quantum Physics, USTC}

\title[An \textsf{achemso} demo]
  {Calibration-free vector magnetometry using nitrogen-vacancy center in diamond integrated with optical vortex beam}

\abbreviations{IR,NMR,UV}

\begin{document}







\begin{abstract}
We report a new method to determine the orientation of individual nitrogen-vacancy (NV) centers in a bulk diamond and use them to realize a calibration-free vector magnetometer with nano-scale resolution. Optical vortex beam is used for optical excitation and scanning the NV center in a [111]-oriented diamond. The scanning fluorescence patterns of NV center with different orientations are completely different. Thus the orientation information of each NV center in the lattice can be known directly without any calibration process. Further, we use three different-oriented NV centers to form a magnetometer and reconstruct the complete vector information of the magnetic field based on the optically detected magnetic resonance(ODMR) technique. Comparing with previous schemes to realize vector magnetometry using NV center, our method is much more efficient and is easily applied in other NV-based quantum sensing applications.
\end{abstract}


\section{Introduction}
Detection of weak magnetic fields with nano-scale resolution is of great importance in fundamental physics, medicine technology and material sciences\cite{degen2017quantum,barry2019sensitivity,kimball2013optical}. Atomic magnetometers\cite{budker2007optical} can detect magnetic field with extremely high sensitivity, but suffer from low spatial resolution. Superconducting quantum interference devices (SQUIDs)\cite{vasyukov2013scanning} can reach nanoscale resolution, but they have a finite size and act as perturbative probes over a narrow temperature range.
Recently, NV center in diamond is emerging as a promising candidate for nano-scale magnetometry and solid-state quantum information processing\cite{DOHERTY2013,Du2015science,Hanson2012register,Degen2019nature,Bing2020APL}. The electron spin in NV center, often with the assistance of proximate nuclear spins, works as a quantum sensor with high sensitivity and long coherence time. It is widely used in room-temperature magnetic sensing and capable of measuring spins in a single-molecular level\cite{grinolds2013nanoscale,maze2008nanoscale,chaudhry2015detecting,wang2015high, glenn2018high, wang2019nanoscale, degen2008scanning, casola2018probing,Hanson2011NJP, Suter2017NJP, chen2019quantum, xu2019dynamically}.

Reconstructing full information of the magnetic field vector (\emph{i.e.}, including magnitude and orientation) is often crucial in applications such as biological magnetic field sensing and the condensed matter physics\cite{le2013optical, schirhagl2014nitrogen, mcguinness2011quantum, casola2018probing}. For single NV center, the basic idea of measuring magnetic field is to detect the frequency shift due to the Zeeman effect by the optically detected magnetic resonance (ODMR) procedure\cite{Jorg2008nature}. In weak magnetic field regime, it measures only the magnetic field along the axis defined by the nitrogen atom and the vacancy in the diamond lattice. To realize a real-time vector magnetometer in a three-dimensional space, one needs to integrate at least three unparalleled scalar magnetometers together.  Because of the $C_{3\nu}$ symmetry of the diamond lattice, there're four kinds of NV centers with different tetrahedral orientations referring to the laboratory frame. Thus it is naturally to form an vector magnetometer with high resolution utilizing the four different-oriented NV centers in a bulk diamond\cite{maertz2010vector,pham2012enhanced}. However, NV center is randomly generated in the diamond, it is not possible to know the position and direction of each NV center in advance. In a vector magnetometry application, we firstly locate the NV centers using a confocal system and then do a calibration process to know exactly the orientation of each NV center.

\begin{figure}[t]
\centering
\includegraphics[width=1.0\textwidth]{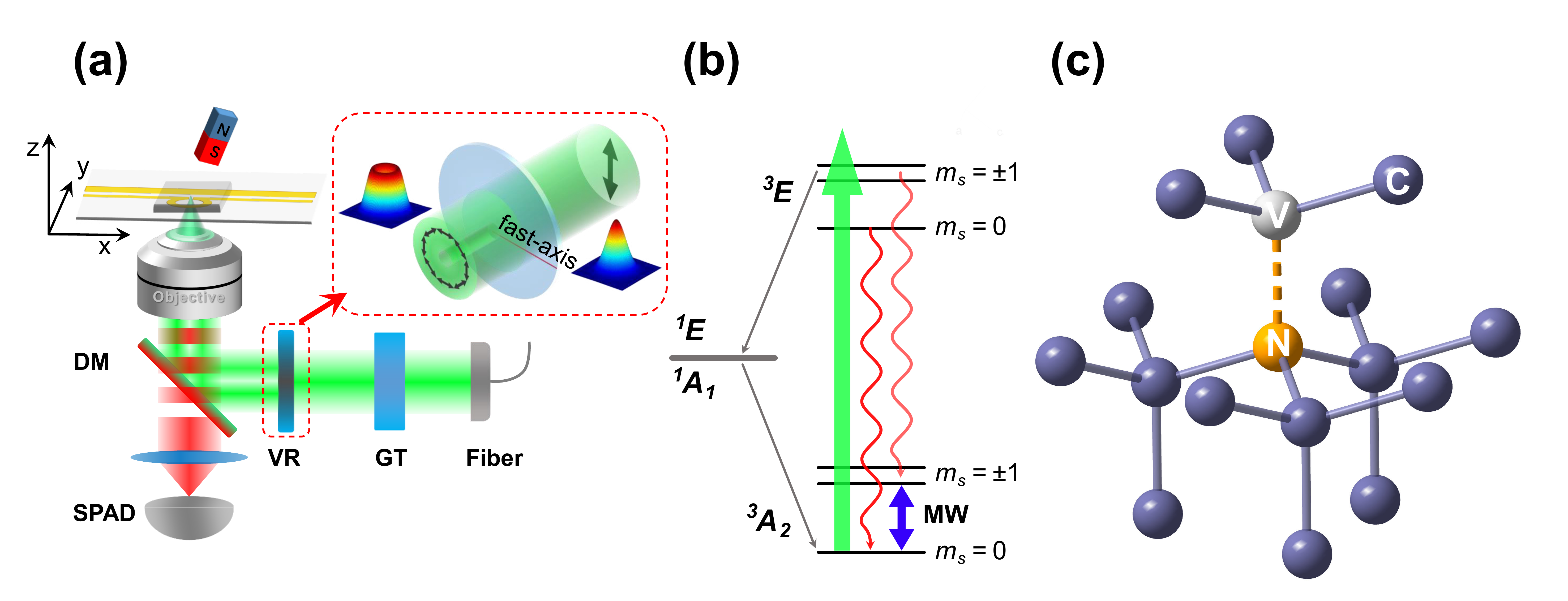}
\caption{\label{Fig 1}(color online). (a) Schematics of the experimental setup with a home-built confocal microscope. The inset shows that the linearly polarized beam is passed through the vortex retarder, which can generate the azimuthally polarized beam. VR: Vortex Retarder; GT: Glan-Taylor Polarizer; DM: dichroic mirror; SPAD: single photon avalanche diode. (b) Scheme of energy levels of the NV center electron spin. Its ground state ($^3A_2$) and excited state ($^3E$) are both spin triplets. The ground state ($^3A_2$) has a zero-splitting (2.87GHz) between the $m_s=0$ and $m_s=\pm1$. The excited state (\emph{$^{3}\text{E}$}) is governed by spin-orbit and spin-spin interactions, split by $1.43$ GHz between $m_s=0$ and $m_s=\pm 1$ states. All spin sub-levels ($m_s=0, \pm 1$) in the excited state exhibit spontaneous decay by photon emission. (c) Diagram of [111]-oriented diamond lattice containing nitrogen(yellow) adjacent to a lattice vacancy(gray).}\label{fig1}
\end{figure}

One conventional way to calibrate the NV magnetometer is to measure the ODMR spectra at given vector magnetic fields generated by a three-dimension electromagnet. By analyzing the ODMR spectra, the orientation of NV center could be figured out in experiment. Recently another optical calibration method is also developed by continuously changing the polarization of the pumping laser\cite{weggler2019determination}. It is because that the photoluminescence of NV center depends on the overlap between the polarization of the laser and the orientation of the NV center\cite{photolumine}. However, both these two calibration methods are time-consuming, preventing a real-time application of vector magnetometry based on NV centers.

In this letter, we report the realization of a calibration-free vector magnetometer with NV centers by using the optical vortex technique. The optical vortex beam with spatially varying electric filed polarization is a fast-developed technique, and is widely used in atomic physics and quantum optics\cite{Zhan:09, Maurer2007NJP, Kimura1995PRL, Zhan2004OE, Huang2012OL, Dorn2003PRL, Cardano2012AO, ye2019experimental}.Azimuthally polarized laser beam, as a kind of vortex beam, is employed to excite the NV centers. The fluorescence pattern of the NV center collected by the confocal microscopic system is dependent on its orientation. Thus we integrate the locating process and the calibration process by scanning the fluorescence of the NV center. We realize this process in a [111]-oriented bulk diamond and further measure the magnetic field vector with standard ODMR procedures using three different-oriented NV centers in the laboratory frame.  This new method opens the way towards a real-time nano-scale vector magnetometer and is generally applicable in other NV-based sensing applications.

\section{Results and discussion}
The spin-Hamiltonian of the NV center can be written as the sum of zero-field splitting term, electron spin and nuclear spin Zeeman splitting terms, the hyper-fine interaction term (hyperfine splitting tensor \textbf{A}), nuclear quadrupole interaction (\emph{Q})\cite{Jorg2008nature,barry2019sensitivity}:
\begin{equation}\label{Hamiltonian}
\hat{H}=D\hat{S}^2_z+g_e{\mu_B}\vec{B}\cdot{\hat{S}}+\underline{\hat{S}}\cdot\textbf{A}\cdot\underline{\hat{I}}+Q\hat{I}^2_z+g_N{\mu_N}\vec{B}\cdot{\hat{I}}
\end{equation}
where \emph{D}=2.87 GHz is zero-field splitting parameter. $\mu_B$ is the Bohr magneton, $\mu_N$ is the nuclear magneton. $g_e$ and $g_N$ are the electron spin and $^{14}N$ nuclear spin g-factor, respectively. In the Hamiltonian, the first term denotes the zero field splitting (ZFS), the second term is the electron spin Zeeman splitting, the third term is the hyper-fine interaction between electron spin and nuclear spin, the fourth term is the nuclear quadrupole splitting ($I>1$), the last term is the nuclear Zeeman splitting. As shown in Fig.\ref{Fig 1}(b), the $m_S=\pm1$ sub-levels of the ground state ($^3A_2$) are degenerate at zero magnetic field. At external magnetic field condition, the $m_S=\pm1$ sub-levels have Zeeman splitting, and the resonance lines between $m_S=0$ and $m_S=\pm1$ are separated around 2.87 GHz. The transitions between the $m_S=0$ and $m_S=\pm1$ spin sub-levels of the ground state ($^3A_2$) can be detected by ODMR.

\begin{figure}[t]
\centering
\includegraphics[width=0.7\textwidth]{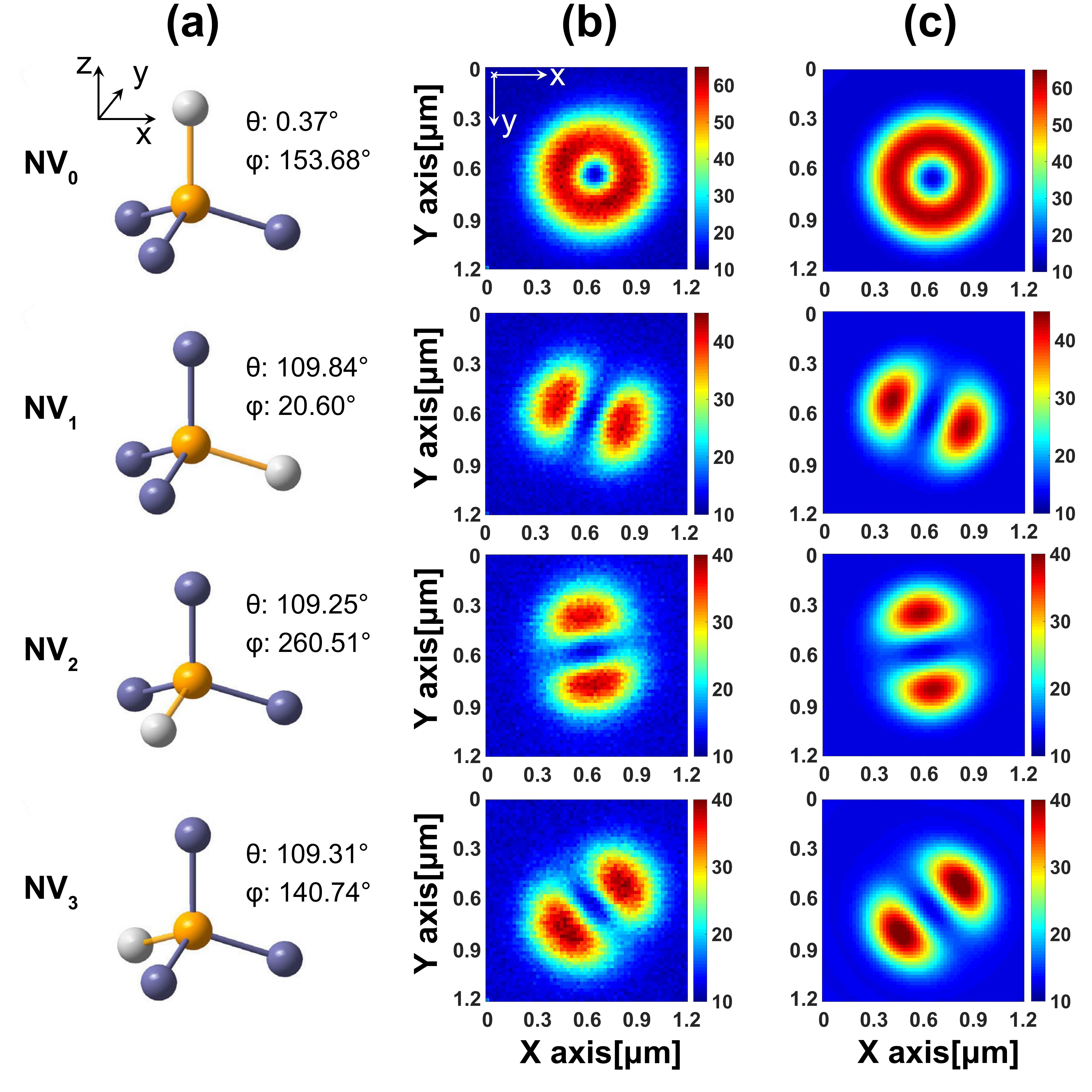}
\caption{\label{Fig 2}(color online). (a) Schematic drawing showing the four NV center orientations for [111]-oriented diamond. The yellow sphere represents nitrogen atom that replaces the original carbon atoms, and the gray sphere represents vacancy. (b) The fluorescence patterns of the four different oriented NV centers. (c) The fitted patterns of four different oriented NV centers. We fit the fluorescence patterns based on pattern matching algorithm by Python and obtain the orientation information of four NV centers. The polar angle $\theta$ and azimuth angle $\varphi$ of four NV centers are (0.37$^\circ$, 153.68$^\circ$), (109.84$^\circ$, 20.60$^\circ$), (109.25$^\circ$, 260.51$^\circ$), and (109.31$^\circ$, 140.74$^\circ$).}\label{fig2}
\end{figure}

In contrast to previous work, we employ the azimuthally polarized beam, which is tightly focused by a high numerical-aperture (\emph{NA}) aplanatic objective lens, to scan the NV centers. The polarization near the focal spot region is inhomogeneous. The electric field vector ($\emph{\textbf{E}(\textbf{r})}$) near focus can be expressed as\cite{Zhan:09}
\begin{equation}\label{azimuthal}
\begin{split}
&E_\varphi(r,z)=2A\int_{0}^{\alpha}\sqrt{cos\theta}(\sin\theta) J_1(kr\sin\theta)e^{ikz\cos\theta}d\theta. \\
\end{split}
\end{equation}
Here $J_n(x)$ is the Bessel function of the first kind with order \emph{n}. $r=\sqrt{x^2+y^2}$ is the position of the focus spot. $k=2\pi/\lambda$ is the wave number of the light in the medium. In the integral, the upper bound $\alpha$ of the polar angle ($\theta$) is determined by the \emph{NA} of the objective lens. \emph{A} is the field strength at the pupil aperture. For azimuthally polarized beam, the total field is transverse and we only consider the azimuthal component of the electric field vector near focus region. Its intensity is an on-axis null and annular intensity distribution at all distance $z$ from the paraxial focus.

The fluorescence intensity distribution is determined by the interaction between the electric field vector and the excitation dipole of the NV center\cite{karedla2015simultaneous, dolan2014complete,Patra2004}. Strain-dependent optical measurement of NV center has indicated that two orthogonal electric dipoles are in the plane perpendicular to the symmetry axis\cite{Epstein2005np}, which are responsible for the emission fluorescence. The fluorescence intensity distribution by scanning NV center can be written as $I(x,y)\propto\left|\emph{\textbf{E}(\textbf{r})}\cdot\bf{\mu_{exc1}}\right|^2+\left|\emph{\textbf{E}(\textbf{r})}\cdot\bf{\mu_{exc2}}\right|^2$, where $\emph{\textbf{E}(\textbf{r})}$ is the position dependent electric field and the $\bf{\mu_{exc1,2}}$ is the excitation dipole vector.

Because of the inhomogeneous transverse polarization components around the focal spot, the scanning confocal image has characteristic intensity pattern that is different from the Gaussian distribution with a linearly polarized beam. The fluorescence pattern with the azimuthally polarized beam is dependent on the three-dimensional orientation of the NV center's excitation dipole. We simulate this process and calculate the photon pattern numerically. In Fig.\ref{fig2}(c), different patterns indicate different oriented NV centers.

The experiment is performed on a purpose-built confocal microscopy system as shown in Fig.\ref{fig1}(a) to address single NV center in a type-IIa, single-crystal synthetic bulk diamond. A collimated 532 \emph{nm} pulse laser beam is passed through the single mode optical fiber and shaped into approximately ideal Gaussian beam. The Gaussian beam is sent through the Glan-Taylor polarizer and produced high-quality linear polarization  beam with an extinction ratio greater than 100,000:1. Such a beam is then sent to the Zero-Order Vortex Half-Wave Retarder(m=1) which can generate azimuthally polarized beam with the doughnut-like feature as shown in Fig.\ref{Fig 1}(a). The generated azimuthally polarized beam is then focused on the NV center using an Olympus oil-immersion microscope objective lens (\emph{NA}=1.40). A piezoelectric transducer (PZT) stage holds the diamond perpendicular to the beam in order to implement the scanning in the transverse \emph{x-y} plane of the sample. The fluorescence photons  ranging from 600 to 800 \emph{nm} are collected by the same objective lens, and detected by using the single photon avalanche diode (SPAD). The microwave signal (generated by Rohde $\&$ Schwarz) is amplified (Mini-Circuits ZHL-42W+) and delivered to an impedance-matched copper slotline (0.1$mm$ gap and an $\Omega$-type ring in the middle) deposited on a coverslip, and finally coupled to the NV center. A small permanent magnet (south pole) in the vicinity of the diamond generates the static magnetic field to be detected.

\begin{figure}[t]
\centering
\includegraphics[width=1.0\textwidth]{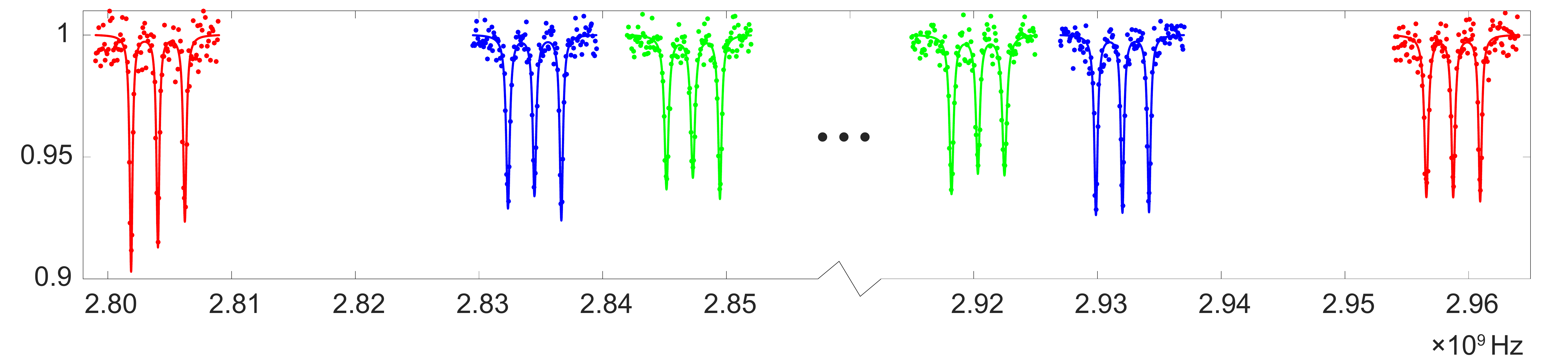}
\caption{\label{Fig 3}(color online). ODMR spectra for different orientation NV centers. In the static magnetic field, different orientation NV centers have anisotropic Zeeman splitting and ODMR spectra. Due to hyperfine interaction with the electron spin and $^{14}N$ nuclear spin, the transitions between $m_s=0$ and $m_s=\pm1$ both have three peaks, respectively. These points and solid lines (red, blue, and green) represent experimental data and data fitting using Lorentzian functions for NV$_1$, NV$_2$, and NV$_3$.}
\end{figure}

\begin{table}[b]
\caption{The location of NV centers, the polar angles $\alpha$ between the static magnetic field and the NV axis, and the magnetic filed magnitude extracted from the ODMR spectra of NV$_1$, NV$_2$, NV$_3$ shown in Fig.\ref{Fig 3}. The location of NV centers in diamond is also shown in scanning confocal image (part 1 of the Supporting Information). }
\label{tab:field}       
 \begin{tabular}{llll}
  \hline\noalign{\smallskip}
     & NV$_1$ & NV$_2$ & NV$_3$  \\
  \noalign{\smallskip}\hline\noalign{\smallskip}
  (x,y) ($\mu$m)  & (16.175, 7.335)& (16.450, 1.920)&(5.346, 7.678)\\
        $\alpha$ ($^{\circ}$) &  117.62 $\pm$ 0.02  &  106.96 $\pm$ 0.02  &  102.55 $\pm$ 0.01 \\
  B (G)     &  59.53 $\pm$ 0.26    &  59.48 $\pm$ 0.35    &  59.56 $\pm$ 0.36  \\
  
  \noalign{\smallskip}\hline
  \end{tabular}
\end{table}

Experimentally, we scan the diamond using the azimuthally polarized beam with the result shown in Fig.\ref{fig2}(b), where four kinds of patterns are all observed and coincident with numerical simulation. To figure out the orientation of NV centers in the laboratory frame, we select four different patterns, which are noted as NV$_0$, NV$_1$, NV$_2$, and NV$_3$. Then we develop an optimization algorithm based on the Nelder-Mead method to determine the angles associated with the scanned patterns. The optimization routine starts from a random state of the angles and minimize the difference between simulation result and experimental pattern by varying the angles iteratively. Note that, there's still a rotational symmetry of $180^{\circ}$ in the azimuth angle, which can be easily broken by using a simple static magnetic field\cite{weggler2019determination}. Fig.\ref{fig2} shows the final result including polar angle $\theta$ and azimuth angle $\varphi$  of the NV centers, where all four patterns match well with experiment. 

\begin{figure}[t]
\centering
\includegraphics[width=0.6\textwidth]{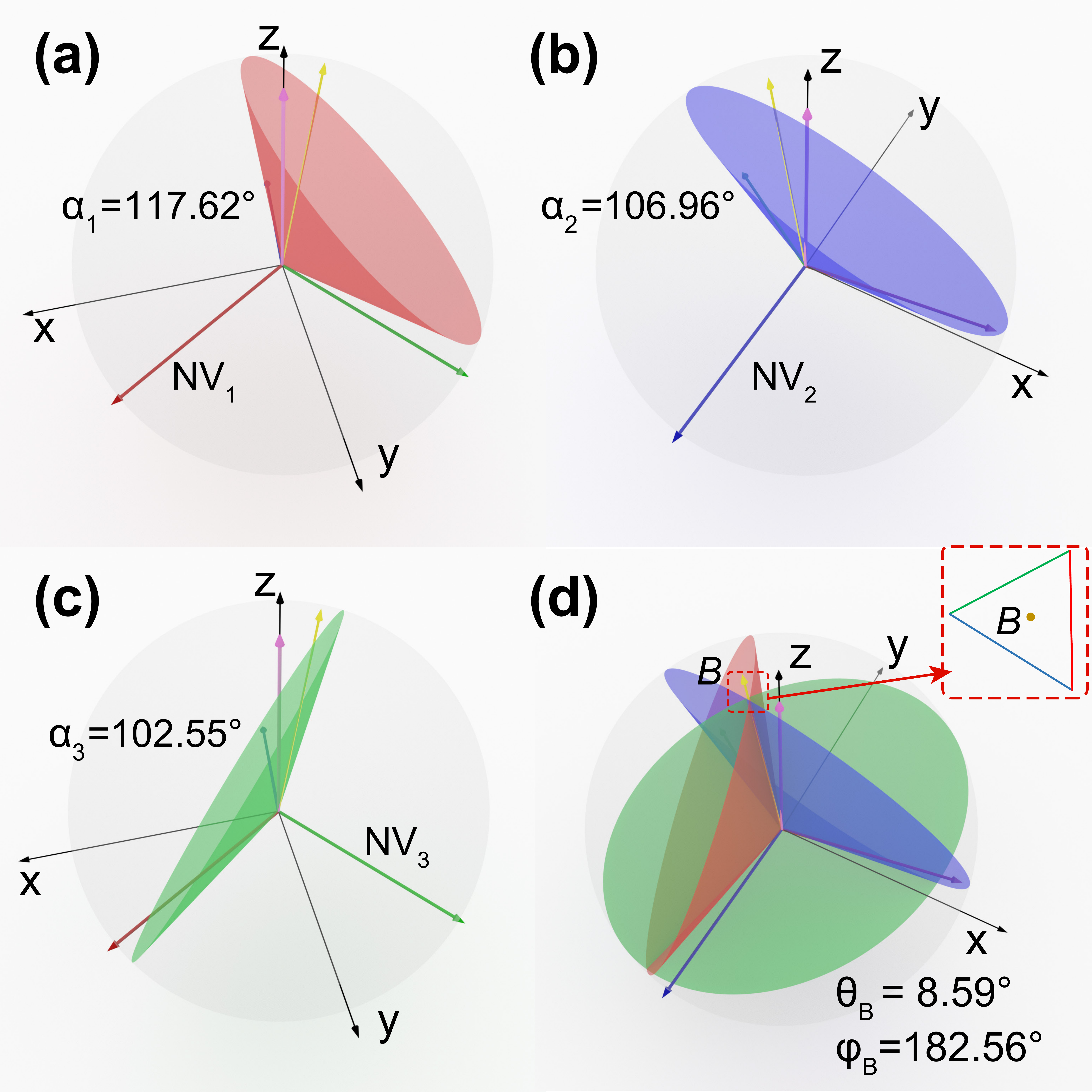}
\caption{\label{Fig 4}(color online). (a)-(c) The possible orientation of the magnetic field. The polar angles $\alpha$ between the magnetic field and the NV$_{1-3}$ center axes are $117.62^{\circ}$, $106.96^{\circ}$, $102.55^{\circ}$ respectively. Which tell the possible magnetic field vector is a cone around the NV center axis. (d) The determination of the magnetic field direction with three different-oriented NV centers. The intersections of three cones form a triangle. The least square method is employed to get the optimal solution, which is $\theta_B=8.59^{\circ},\varphi_B=182.56^{\circ}$.}
\end{figure}

With all the obervations, we then proceed to demonstrate the vector magnetometer with three different-oriented NV centers. And the full vector information of the field could be reconstructed from the individual measurement results and the NV center orientation. In each measurement, the magnitude of magnetic field, the polar angle $\alpha$ between the magnetic field and the NV center axis are obtained by the standard ODMR procedure.

The ODMR spectra lines are usually power broadened by laser and microwave (MW) radiation. We employ the normal Gaussian beam to excite the NV centers and get the ODMR spectra. To obtain the high resolution ODMR spectra, we replace the CW (continuous wave) ODMR process with the pulsed ODMR process using pulsed laser ($\sim$400 ns) and MW field (electron spin $\pi$-pulse). By sweeping MW field frequency, we can observe the electron spin transitions from $m_S=0$ to $m_S=\pm1$ in ground state ($^3A_2$), which are both three peaks owing to the hyper-fine interaction between electron spin and $^{14}N$ nuclear spin ($I_{^{14}N}=1$) based on Eq.\ref{Hamiltonian}. As shown in Fig.\ref{Fig 3}, the middle peaks ($\omega_1$ and $\omega_2$) of the three peaks are the transition frequencies of $\left |m_s=0, m_I=0\right\rangle$ $\rightarrow$ $\left |m_s=-1, m_I=0\right\rangle$ and $\left |m_s=0, m_I=0\right\rangle$ $\rightarrow$ $\left |m_s=+1, m_I=0\right\rangle$, respectively. For magnetic field in arbitrary direction, the transition frequencies can be selected to calculate the magnitude according to $B=\sqrt{(1/3)(\omega_1^2+\omega_2^2-\omega_1\omega_2-D^2)}/(g_e\mu_B/\hbar)$ and the polar angle $\alpha\in[0,\pi]$ expressed by $\alpha=arccos(\pm\sqrt{\frac{(2\omega_1-\omega_2-D)(\omega_1-2\omega_2+D)(\omega_1+\omega_2+D)}{[9D(\omega_1^2-\omega_1\omega_2+\omega_2^2-D^2)]}})$\cite{Jorg2008nature}. Based on the ODMR spectra, we only can get the polar angle ${\alpha}$ between the magnetic field and the NV axis. But the polar angle ${\alpha}$ only tells that the possible orientation of the magnetic field is a cone around the NV center axis as shown in Fig.\ref{Fig 4}(a)-(c). 

To completely obtain the information about the magnetic field orientation, 
we need three different orientation NV centers at least. Here, we select NV$_1$, NV$_2$ and NV$_3$ as magnetic sensors.
The fitted results of the polar angle and magnitude for NV$_1$, NV$_2$ and NV$_3$ are shown in Table \ref{tab:field}. In theory, the cones around three NV center axes have an intersection, which is the absolute orientation of the magnetic field in laboratory coordinate system as shown in Fig.\ref{Fig 4}. Experimentally, three cones can't strictly intersect at a point and have deviation in the overlap. The deviation can be from strain of lattice and temperature fluctuation during measurement process. The intersections of three cones form a triangle in unit sphere as shown in the inset of Fig.4(d). To evaluate the direction of the magnetic field, we employ the least square method to get the optimal solution (see Supporting Information). The direction of the magnetic field is $\theta_B=8.59^{\circ},\varphi_B=182.56^{\circ}$ with error less than $0.63^{\circ}$.


\section{Conclusion}
In conclusion, we propose and demonstrate an efficient way to reconstruct vector information of magnetic field with NV centers in diamond. The optical vortex beam is azimuthally polarized beam which induces an orientation-dependent image pattern when scanning NV centers in a confocal system. With the vortex beam, the orientations of the NV centers in a bulk diamond could be directly determined from the scanning image. Combining with the ODMR spectra, the complete information of the magnetic field including the magnitude and orientation could be reconstructed. Our works provides a calibration-free nano-scale vector magnetometry and can be easily used in NV-based magnetic field sensing and imaging.

\begin{acknowledgement}
The authors are grateful to Heng Shen, FengJian Jiang and Yong Zhou for fruitful discussions. This work is supported by the National Key R\&D Program of China (Grants No. 2018YFA0306600, No. 2018YFF01012505 and No. 2018YFF01012500), the National Natural Science Foundation of China (Grants No. 11604069, No. 61805064, No. 11775209, No. 81788101, No. 11761131011  and No. 11904070), the Fundamental Research Funds for the Central Universities, the CAS (Grants No. GJJSTD20170001, No. QYZDY-SSW-SLH004),
the Anhui Initiative in Quantum Information Technologies (Grant No. AHY050000).
\end{acknowledgement}




\bibliography{achemso-demo}

\end{document}